\documentclass[aps,]{revtex4}%
\usepackage{amsfonts}
\usepackage{amsmath}
\usepackage{amssymb}
\usepackage{graphicx}%
\begin{document}
\title[Thermoelectric efficiency]{Thermoelectric efficiency of heterogeneous media at low temperatures}
\author{S.A. Ktitorov, V.K. Zaitsev, M.I. Fedorov}
\affiliation{A.F. Ioffe Physico-Technical Institute of the Russian Academy of Sciences,
Polytechnicheskaja str. 26, 194021 St. Petersburg, Russia}
\author{}
\affiliation{}
\author{}
\affiliation{}

\begin{abstract}
A quantum limit of the thermoelectric efficiency for heterogeneous media with
weak links is established with a use of the Landauer-type formulae.

\end{abstract}
\maketitle

%\preprint{HEP/123-qed}

%\volumeyear{ }
%\volumenumber{ }
%\issuenumber{ }
%\eid{ }
%\date{}
%\received[Received text]{}

%\revised[Revised text]{}

%\accepted[Accepted text]{}

%\published[Published text]{}

%\startpage{ }
%\endpage{ }
%\tableofcontents

%\pacs{PACS number}

\section{Introduction}

As is well known, the kinetic parameters of semiconductors and metals such as
conductivity $\sigma$, thermoelectric power $S$ and thermoconductivity
$\kappa$ are determined by a crystallographic and atomic structure of the
material and are, therefore, very specific for an every given material
\cite{Ans}. However, there is a class of conductors that have low-temperature
kinetic properties, which are not specifically related to a concrete
scattering mechanism and to other nonuniversal factors. We will show here that
the thermoelectric efficiency $zT=\frac{S^{2}\sigma T}{\kappa}$ of these
structures at low enough temperatures can be expressed in terms of the
fundamental constants and of the system geometry.

\section{Conducting structures with weak links}

Let us consider a polycrystal with point contacts between crystallytes. These
point contatcs are really small bridges or weak links between the
crystallites. We assume that within the actual temperature range, all
quasiparticles (charge carriers, phonons, etc) are almost thermalized in
crystallites and their distribution functions insignificantly differ from the
equilibrium ones. Large volume of crystallites in comparison with bridges
connecting them ensures effective thermalization. Notice, that by our opinion,
it is not even necessary for strong inequality $D\gg l$ (with $l$ being the
mean free path) to be fulfilled because nonspherical crystallites can play a
role of three-dimensional Sinai's billiards \cite{Zas}\ for most of the
quasiparticles. In other words, large time delay in propagation of
quasiparticles in result of multiple reflections on the crystallite boundary
leads to effective thermalization, if the total path through the crystallite
including multiple reflections by the crystallite boundaries exceeds the mean
free path. Similar processes can occur in conducting composites based on
opals, porous glasses etc. Rather narrow windows in the spectrum are formed by
excitations, which avoided multiple reflections because of a spatial coherence
resonance. Such windows of coherence can be actual only at very low
temperatures. Large cross section of a crystallite makes the electric current
and heat flux densities extremely small there. Thus, we can consider the
system as a network of electric and thermal resistors and thermoelements (weak
links) connecting nodes (crystallites), which can be characterized by their
electric potentials and temperatures .Then the transport problem is reduced to
the following two steps: (i) calculation of the electric and thermal
resistance and of the thermoelectric power of a single weak link; (ii)
calculation of integral characteristics of the network. The latter is
extremely simple in the case of a regular network, but can be rather
complicated in the irregular case. In the case of a strong irregularity
(severed connections, in particular), the percolation theory can be applied.

\section{Transport properties of a weak link}

Let us consider a bridge connecting two crystallites as a figure of revolution
only slightly different from a cylinder. Phonon and electronic contributions
have to be considered separately.

Long wavelength excitations propagate without feeling of possible disorder
that allows us to use the quasiparticle picture at low temperatures. Size
quantization of excitation spectra in the weak link permits us to consider it
as a waveguide for excitations that establishes a lower limit for propagating
waves using the relation:%

\begin{equation}
\beta_{n}=\sqrt{k^{2}-\varkappa_{n}^{2}}, \label{waveguide}%
\end{equation}
where $k$ is the wave vector value, $\beta_{n}$ is the wave vector component
along the waveguide axis, $\varkappa_{n}$ is the membrane eigenvalue, the
lowest one is of order $d^{-1}$. Simple estimates show that these wavelengths
are related for acoustic phonons and electrons, respectively, to energies of
order 1 $K$ and 10 $K$. Therefore, only a few waveguide modes participate in
transport processes within this temperature range, but their number fast
increases with energy. This conclusion is valid if we assume that actual
excitations feel the weak link as a regular waveguide that can be correct only
for the most low-lying excitations with large wave length. When the energy of
an excitation increases, the bridge becomes more irregular for waves. The
transmission coefficient decreases catastrophically except for a few (if they
exist) special energy values related to possible resonances in irregular
interfaces. Besides, there exist gapless surface modes.

The one-dimensional electronic ballistic transport in the presence of the
temperature and electrochemical potential is described by the equation
\cite{Landauer}:%

\begin{equation}
J=-\frac{2e^{2}}{2\pi\hslash}\sum_{\nu}\int_{0}^{\infty}dE\frac{dn_{F}(E)}%
{dE}t_{\nu}^{el}(E)\left[  \frac{E-\mu}{T}\Delta T+\Delta\mu\right]  ,
\label{electrcurr}%
\end{equation}
where $J$ is the electric current through the link, $n_{F}$ is the Fermi-Dirac
distribution, $t_{\nu}^{el}\left(  E\right)  $ is the electron transmission
coefficient, $\mu$ and $\Delta\mu$ are respectively the electrochemical
potential and its drop across the weak link, $\Delta T$ is the temperature
difference between the adjacent grains. The energy flow $Q=Q^{ph}+Q^{el}$
comprises contributions both from the electronic and phonon subsystems%

\begin{align}
Q^{ph}  &  =\sum_{\nu}\int_{0}^{\infty}\frac{dk}{2\pi}\hslash\omega_{\nu}%
\frac{\partial\omega_{\nu}}{\partial k}t_{\nu}^{ph}\left(  n_{R}^{ph}%
-n_{L}^{ph}\right)  ,\label{phononheat}\\
Q^{el}  &  =\sum_{\nu}\int_{0}^{\infty}\frac{dk}{2\pi}\epsilon_{\nu}\left(
k\right)  \frac{\partial\epsilon_{\nu}}{\partial k}t_{\nu}^{el}\left(
n_{R}^{el}-n_{L}^{el}\right)  \label{electronheat}%
\end{align}
where $\omega_{\nu}\left(  k\right)  $ is the $\nu$-th acoustic phonon branch
frequency, $n_{R}^{ph}$ and $n_{L}^{ph}$ are respectively the quasiequilibrium
phonon distribution functions in the bulk of the right and left crystallites;
$\epsilon_{\nu}\left(  k\right)  $ is the $\nu$-th branch of the electronic
spectrum in the link, .$n_{R}^{el}$ and $n_{L}^{el}$ are corresponding
electronic distribution functions.

\section{Kinetic properties of the network}

Within the approximation of locally equilibrium crystallites, the electric
potential and temperature distributions are governed by a finite-difference
equations. In the steady state, the electric potential and temperature
distributions satisfy the time-independent finite-difference electric and heat
transport equations: the ''Kirhgoff law'' for currents entering into the
$\mathbf{n}$-th site%

\begin{align}
\sum_{\left\langle \mathbf{m}\right\rangle }J_{\mathbf{nm}}  &
=0,\label{kirh1}\\
\sum_{\langle\mathbf{m}\rangle}Q_{\mathbf{nm}}  &  =0 \label{kirh2}%
\end{align}
and the Ohm-Fourier-Peltier-Zeebeck law for the electric $J_{\mathbf{mn}\text{
}}$and heat $Q_{\mathbf{nm}}$ currents between the sites $\mathbf{m}$ and
$\mathbf{n}$\bigskip%

\begin{align}
Q_{\mathbf{nm}}  &  =-G_{\mathbf{nm}}(T_{\mathbf{n}}-T_{\mathbf{m}}%
)+\Pi_{\mathbf{nm}}J_{\mathbf{nm}}\nonumber\\
-(V_{\mathbf{n}}-V_{\mathbf{m}})  &  =R_{\mathbf{nm}}J_{\mathbf{nm}%
}+S_{\mathbf{nm}}(T_{\mathbf{n}}-T_{\mathbf{m}}), \label{onzag}%
\end{align}
where $\mathbf{n}$ stands for the crystallite number vector with components
$n_{x}$, $n_{y}$, $n_{z}$; $G_{\mathbf{nm}}$ is the thermal conductance of the
weak link connecting crystalites $\mathbf{m}$ and $\mathbf{n}$, $G_{mn}\neq0$
if and only if $\mathbf{m}$ and $\mathbf{n}$ differ by a unit superlattice
vector $\mathbf{e}$.; $\left\langle \mathbf{m}\right\rangle $ means summing
over the nearest neighbours of the given node $\mathbf{n}$., $R_{\mathbf{nm}}$
is the electric resistance of the link, $V_{\mathbf{n}}$ is the potential of
the n-th crystallite, the thermopower $S_{\mathbf{nm}}$ and the Peltier
coefficient $\Pi_{\mathbf{nm}}$ are related by the Kelvin-Onsager symmetry law
$\Pi_{\mathbf{nm}}=S_{\mathbf{nm}}(T_{\mathbf{n}}+T_{\mathbf{m}})/2$,
$Q_{\mathbf{nm}}$ is the energy flow through the link, $J_{\mathbf{nm}}$ is
the electric current through the link. Now the theoretical analysis of the
heat transport problem for the polycrystal is reduced to two different ones:
the first problem is to study the electric and heat transport through the weak
link and the second is to solve the network equations for a given distribution
of links parameters. The former can be considered using different methods
depending on the temperature range, the weak link geometry, and the
microscopic properties of the material, while the latter is determined
exclusively by the system geometry and a distribution of parameters. For a
regular superlattice we have $G_{\mathbf{n,n+e}}=G$ for all $\mathbf{n}$ and
$\mathbf{m}$, and the effective thermoconductivity of the sample reads%

\begin{equation}
\varkappa_{eff}=G/A \label{thermcond}%
\end{equation}
where $A$ is the superlattice spatial period. A calculation of the effective
thermoconductivity in the case of a random distribution of $G_{\mathbf{mn}}$
can be very sophisticated problem mathematically similar to the electron
transport in irregular conductors, but we concentrate our attention here on
the weak link thermal resistance assuming that the superlattice is regular.

\section{Efficiency}

Assuming the structure to be regular, we calculate the thermoelectric
efficiency of a single link. The thermoelectric efficiency can be expressed in
terms of the single link transport coefficients:%

\begin{equation}
zT=\frac{S^{2}G_{el}T}{G_{th}} \label{zT}%
\end{equation}
where $S$, $G_{el}$, and $G_{th}$ are respectively the thermopower, the
electric and thermal conductances of the single link, which are given by the
formulae following from \ref{electrcurr}, ;%

\begin{equation}
S=\frac{k_{B}}{e}\sum_{\nu}\int_{o}^{\infty}dE\cdot t_{\nu}^{el}\frac{dn^{el}%
}{dE}\frac{E-\mu}{k_{B}T}\cdot\left[  \int_{0}^{\infty}dE\cdot t_{\nu}%
^{el}\frac{dn^{el}}{dE}\right]  ^{-1}, \label{thermpower}%
\end{equation}

\begin{equation}
G^{el}=\frac{2e^{2}}{2\pi\hbar}\sum_{\nu}\int_{0}^{\infty}dE\frac{dn_{F}%
(E)}{dE}t_{\nu}^{el}(E), \label{elcond}%
\end{equation}

\begin{equation}
G^{ph}=\frac{\pi k_{B}^{2}T}{6\hbar}\sum_{\nu}\int_{0}^{\infty}dx\frac
{3x\exp\left(  x\right)  }{\left[  \exp\left(  x\right)  -1\right]  ^{2}%
}t_{\nu}^{ph}\left(  x\right)  \label{phonconduct}%
\end{equation}

Notice that the group velocity $\frac{\partial\epsilon}{\partial k}$ and the
one-dimensional density of states $1/\left(  \frac{\partial\epsilon}{\partial
k}\right)  $ cancele one another in the one-dimensional transport. Thermal
conductance is extremely sensitive to the real geometry of the link that can
be seen from the effective dimension of the transport and the behaviour of the
transmission coefficient. In the three-dimensional diffraction regime it reads:%

\begin{equation}
G^{ph}=\frac{\pi^{2}aT^{3}}{30\hbar^{3}} \label{phon3d}%
\end{equation}
In the one-dimensional case it can be written as%

\begin{equation}
G^{ph}=\frac{k_{B}^{2}\pi^{2}}{6\pi\hbar}TN_{ph}, \label{phon1d}%
\end{equation}
where $a$ is the link cross-section area, $N^{ph}$ is the number of the
gapless phonon modes.

We can write within this approximation for the electric conductance :%

\begin{equation}
G^{el}=\frac{e^{2}}{\pi\hbar}N^{el}, \label{electrcondapprox}%
\end{equation}
While the conductance and the thermoconductance have a universal form, which
is very not sensitive to details of the weak link geometry,%

\begin{equation}
S=\frac{k_{B}}{e}\frac{\ln2}{N^{el}+\frac{1}{2}}. \label{thermpowerappr}%
\end{equation}

Substituting these formulae into \ref{zT}, we obtain for the thermoelectric
efficiency quantum limit at the low-temperature regime%

\begin{equation}
zT=\frac{6\left(  \ln2\right)  ^{2}}{\pi\left(  N^{el}+\frac{1}{2}\right)
^{2}\left(  1+\frac{N^{ph}}{N^{el}}\right)  } \label{fin}%
\end{equation}

This means that one has to minimize the number of modes involved into the
transport through the weak link in order to rise the thermoelectric efficiency.


\begin{thebibliography}{9}                                                                                                %
\bibitem {Ans}A.I. Anselm, \textit{Introduction to the semiconductor theory,
}Nauka,\textit{ }Moscow, 1978.

\bibitem {Zas}G.M. Zaslavskii, \textit{Stochasticity of dynamic systems,
}Nauka, Moscow, 1984.

\bibitem {Landauer}R. Landauer, IBM J. Res. Dev. \textbf{1}, 223 (1957).
\end{thebibliography}
\end{document}